# Battery Valuation and Management for Battery Swapping Station with an Intertemporal Framework

Xinjiang Chen[1], Yu Yang[2], Jianxiao Wang[3], Jie Song[1], Guannan He[1]*

**Abstract**

Battery swapping as a business model for battery energy storage (BES) has great potential in future integrated low-carbon energy and transportation systems. However, frequent battery swapping will inevitably accelerate battery degradation and shorten the battery life accordingly. To model the tradeoff of BES use between energy and transportation applications coupled by battery swapping, we develop a life-cycle decision model that coordinates battery charging and swapping. This model is derived based on an improved intertemporal decision framework, in which the optimal marginal degradation cost (MDC) of BES is determined to maximize the BES benefit across time and application. The proposed framework and model are applied to manage a battery swapping station that simultaneously provides battery swapping services to electric vehicle customers and provides flexibility service to the power grid, including energy arbitrage and reserve. The case study shows that while the end of the physical life of BES occurs faster with battery swapping, the economic life becomes considerably longer. The results also reveal that the optimal MDC depends on the battery values in each application, and we analyze how the battery swapping price affects the optimal MDC and battery life. The proposed framework and model can also provide decision support for on-demand BES service, such as battery trading, renting and secondary use.

[1] Xinjiang Chen, Jie Song and Guannan He are with the College of Engineering, Peking University, Beijing 100091, China.
[2] Yu Yang is with the Department of Industrial and Systems Engineering, University of Florida, Gainesville, FL 116595, USA.
[3] Jianxiao Wang is with the National Engineering Laboratory for Big Data Analysis and Applications, Peking University, Beijing 100091, China
* Corresponding author: Guannan He (e-mail: gnhe@pku.edu.cn)

| Nomenclature | | | |
|---|---|---|---|
| Indices and sets | | Variables and functions | |
| $h$ | indices for time periods, typically an hour | $LB$ | life-cycle objective function of energy storage system |
| $t$ | indices for time periods, typically a day | $LB^*$ | optimal life-cycle objective of energy storage system |
| $T_{life}$ | a set of long-term time periods, typically a life-cycle of a battery | $SB_t$ | short-term objective function of energy storage system |
| Parameters | | $SB_t^*$ | optimal short-term objective of energy storage system |
| $\rho \in [0,1)$ | self-discharge rate | $REV_t$ | market revenue of storage in time period $t$, $ |
| $\eta \in (0,1]$ | charging/discharging efficiency | $C_t^{DEG}$ | total degradation cost of batteries in time period $t$, $ |
| $q_t$ | calendar degradation of battery in time period $t$, MWh-throughput | $C_t^{BS}$ | total labor cost of battery swapping in time period $t$, $ |
| $D$ | total degradation/usage before the life of storage ends, MWh-throughput | $\kappa(t)$ | the year number for time $t$ from the beginning of the battery project |
| $\Delta t$ | time horizon of short-term scheduling | $\mu$ | life-cycle marginal degradation cost (MDC) for storage, $/MWh-throughput |
| $\Lambda_h$ | locational marginal price (LMP) during time $h$, $/MWh | $\mu_t$ | adjusted MDC (AMDC) for storage in time period $t$, $/MWh-throughput |
| $\Lambda_t^{rep}$ | battery swapping price in time period $t$, $/MWh | $d_t$ | degradation of storage in time period $t$, MWh-throughput |
| $c^{bs}$ | labor cost of battery swapping in time period $h$, $/h | $e_h^{cha}$ | amount of energy charged for the storage during time $h$, MWh |
| $E^{max}$ | energy capacity of storage, MWh | $e_h^{dis}$ | amount of energy discharged for the storage during time $h$, MWh |
| $E_h^{max}$ | maximal amount of energy that can be charged/discharged in time period $h$, MWh | $r_h$ | amount of power discharged by replacing batteries during time $h$ in storage, MWh |
| $E_t^{rep}$ | maximal amount of energy that can be swapped in time period $t$, MWh | $E_h$ | energy stored in storage at the end of time period $h$, MWh |
| $\delta_t$ | discounting factor in time period $t$ | $\mathbf{P}_t$ | schedules of battery charging, discharging, and swapping during time $t$, |
| $r$ | discounting rate | $\mathbf{F}$ | feasible operating set of energy storage system, $\mathbf{P}_t \in \mathbf{F}$ |

# 1 Introduction

A high proportion of renewable energy penetration will pose a great challenge to the safe operation of the power system because of its intermittency and volatility. To handle the above challenge, flexible resources, such as battery energy storage (BES), are called to integrate renewable energy. As a key technology for integrating renewable energy, BES is expected to play a crucial role in shaping a clean, sustainable, and low-carbon energy and transportation systems [1][2][3]. In addition to renewable energy integration, BES is also widely applied in energy arbitrage [4], peak shaving [5], frequency regulation [6] and voltage support [7] in power systems.

More recently, battery swapping, an ongoing business model of BES, has received much attention, especially in China, because of its substantial energy arbitrage capability and numerous commercial applications (i.e., battery trading, renting and secondary use [4][8][9]). Specifically, the sales volume of battery swapping vehicles is more than 160,000 with a year-by-year increase of 162% and a market penetration rate of 4.6% in China in 2021, while the total number of battery swapping stations is more than 1,600. It is expected that by 2025, the sales volume of battery swapping vehicles and the number of battery swapping stations in China will exceed 1.92 million and 30,000, respectively. Note that, compared with the charging mode, the deployment of the battery swapping mode is more flexible and requires less charging time (a few minutes). Additionally, the batteries stored in the battery swapping stations can also be used to provide energy services to grids, such as energy arbitrage and reserves, as a secondary application.

Battery degradation has been the major concern for vehicle-to-grid (V2G) [10][11], as have batteries at battery swapping stations. Battery degradation depends on many factors, such as the charge/discharge rate, depth of discharge (DOD), state of charge (SOC), and temperature, which affect the benefits of BES [12][13]. Previous studies on battery degradation models can be classified into physics-based models (PBMs) [14][15], equivalent circuit models (ECMs) [16], machine learning models (MLs) [17][18], empirical models (EMs) [19], and semiempirical models (SEMs) [20]. Most of the above battery degradation models have been integrated into optimization models of power grid applications, and the most conventional approach is regrading battery degradation as a constraint to limit the maximum amount of battery degradation during system operation [21].

However, we are more concerned about how to estimate the battery degradation cost and how battery degradation affects the operating benefits on a life-cycle time scale. He *et al.* [22][23] defined the battery degradation cost as an opportunity cost and developed an intertemporal decision framework to estimate the optimal life-cycle marginal degradation cost (MDC). In addition, He *et al.* [4][24] integrated the battery degradation cost into the objective function of the optimization models and constructed the decision models of the power system and BES that both maximize the benefits. Xu *et al.* [25] proposed a dynamic evaluation framework to determine the financial worth of batteries based on their state-of-life in power grid applications and used the proposed framework to estimate the value of second-life batteries. Note that the above decision models of BES only involve cycle degradation and calendar degradation caused by battery charging/discharging.

With regard to battery swapping, the existing studies mainly focus on the location optimization of battery swapping stations [26][27][28], battery charging [29][30][31][32] and swapping [33][34][35] optimization of electric vehicles (EVs). Lin *et al.* [27] predicted the spatiotemporal demands of battery swapping based on the Monte Carlo algorithm and used the tabu search and genetic algorithm to optimize the locations of battery swapping stations. The results show that optimized locations of battery swapping stations can considerably reduce land rentals. You *et al.* proposed two scheduling methods for the centralized [33] and distributed [34] scheduling problems of battery swapping. In the centralized scheduling problem, the agents involving distribution grids, battery swapping stations, and EVs are managed centrally by the operation center, while the agents can make autonomous decisions in the distributed scheduling problem. To solve the above problems efficiently, You *et al.* [33] developed an exact algorithm named generalized Benders decomposition to obtain the optimal centralized solutions and proposed two quasi-optimal algorithms based on multipliers and dual decomposition to solve the distributed scheduling problem [34]. Li *et al.* [35] studied the day-ahead scheduling problem of EVs with battery charging and swapping coordination, proposed a bilevel model and developed a heuristic based on the alternating direction method to estimate the integer solution of the model. The results show that the proposed model and algorithm can effectively improve the quality of battery swapping services.

Intuitively, battery swapping would inevitably lead to frequent battery charging/discharging, and thus, account for rapid battery degradation, which would further affect the life and benefits of BES. Additionally, the battery swapping prices also play a crucial role in the revenue of the BES operation. Therefore, to accurately evaluate the economics of BES with battery swapping on a life-cycle time scale, the cycle degradation caused by battery charging/swapping, calendar degradation, and the battery swapping price must be factored into the operating model of BES. To the best of our knowledge, no study has focused on the life-cycle MDC and battery swapping price optimization for a commercial/industrial battery swapping station.

In this context, we improved the intertemporal decision framework for the valuation and management of BES with coupled energy and transportation applications. We propose a life-cycle decision model for simultaneous battery charging and swapping considering battery degradation. The model is used to evaluate the trade-off of battery use between energy and transportation applications. The major contributions of this study are as follows.

1) We propose an improved intertemporal decision framework that is suitable for battery energy storage systems, battery swapping stations and EVs to estimate the optimal degradation cost caused by battery charging, discharging and swapping and simultaneously determine the optimal battery swapping prices of battery swapping stations.

2) We develop a life-cycle decision model for battery swapping stations that integrates short-term scheduling and long-term decisions based on the improved intertemporal decision framework. The life-cycle decision model characterizes and optimizes battery charging, discharging, and swapping across energy and transportation applications.

3) We use the above decision framework and model in a commercial/industrial battery swapping station and reveal the effect of battery swapping on the economic and physical life of batteries and how the MDC, battery swapping prices, and electricity prices interact with each other.

The remainder of this paper is organized as follows. In Section 2, we propose the improved intertemporal decision framework and life-cycle decision model formulation for battery swapping stations. Section 3 and Section 4 are devoted to the experimental results and discussions, respectively. Section 5 contains some concluding remarks.

## 2 Methods

### 2.1 An improved intertemporal decision framework

BES owners and operators should optimize battery usage given limited battery life and fluctuating electricity prices. For example, the operators should decide whether to use, how much to use, or which batteries to use for different periods. In this sense, to more accurately evaluate and improve the life-cycle benefits of BES with battery charging and swapping coordination, battery degradation should be incorporated into the life-cycle decision model to achieve the trade-off between short-term utilization and long-term battery life of batteries.

As shown in Fig. 1, we develop an improved intertemporal decision framework for BES from [22]. In the input of the decision framework, a life-cycle decision model that integrates the short-term scheduling model is proposed to simulate the operation of BES and maximize the long-term (life-cycle) benefits by aggregating the short-term benefits. Specifically, given price forecasts (e.g., locational marginal prices (LMPs), battery swapping prices and prices in ancillary service markets) and BES profiles (i.e., SOC and SOH of batteries), the daily schedules of battery charging, discharging, and swapping outputs by the short-term scheduling model. Sequentially, we estimate the mid-term (monthly or annual) and life-cycle benefits of BES by aggregating short-term (daily) and mid-term benefits before the battery life ends (the total degradation/usage of battery). In doing so, we can obtain the BES life-cycle decision.

With regard to the degradation cost of BES, we define the life-cycle MDC of BES based on the loss of future opportunity (i.e., MDC reflects the future/long-term use value of BES) and dynamically update the short-term MDC by the adjusted MDC (AMDC). That is, the short-term MDC (i.e., AMDC) can be computed by the life-cycle MDC. Here we also use the decision framework to analyze how the dynamic battery swapping prices affect the life-cycle benefits of BES. Specifically, we determine the optimal battery swapping price given the predictable market (i.e., the demand-price curve). Meanwhile, we consider the price elasticity of demand for battery swapping and determine the optimal battery swapping price and the corresponding life-cycle MDC for different demand-price curves in the market.

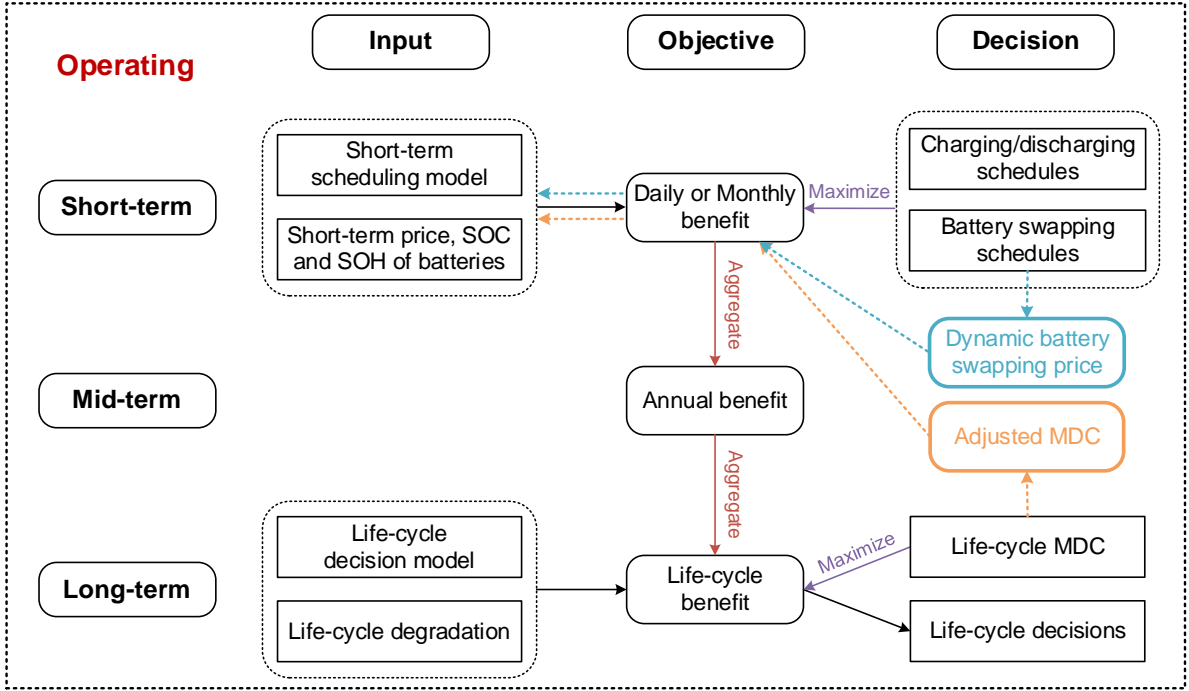

Fig. 1 An improved intertemporal decision framework for BES

## 2.2 A life-cycle decision model for BES

Here we propose a life-cycle decision model for BES with battery swapping to evaluate the cash flow over its life cycle. The model integrates short-term scheduling and long-term decisions for BES. The mathematic formulations of the life-cycle decision model are as follows:

$$LB^* = \max_{\mu} LB = \max_{\mu} \sum_{t \in T_{life}} \delta_t SB_t^*(\mu) \tag{1}$$

$$\text{s.t.} \quad \sum_{t \in T_{life}} d_t(\mu) \leq D \tag{2}$$

$$d_t(t) \geq q_t \tag{3}$$

$$SB_t^* = \max_{\mathbf{P}_t \in \mathbf{F}} SB_t(\mathbf{P}_t) \tag{4}$$

where $LB^*$ and $SB_t^*$ represent the optimal long-term and short-term benefits, respectively. The long-term objective of the BES ($LB$) shown in Equation (1) is to maximize the benefits, and this can be computed by aggregating the discounting short-term benefits $\delta_t SB_t^*$. $\delta_t$ is a typical exponential discounting factor, which can be computed by $\delta_t = (1+r)^{-\kappa(t)}$, where $r$ is the discounting rate and $\kappa(t)$ indicates the year number for time $t$ from the beginning of the battery project. Equation (2) restricts that the degradation and BES usage during the operation periods is no more than the total degradation and usage before the battery life ends. Equation (3) guarantees that the degradation and usage during the operation periods is no less than the calendar degradation of BES. Equation (4) presents the short-term objective of a BES operation, which can be formulated as follows:

$$SB_t^* = \max_{\mathbf{P}_t \in \mathbf{F}} SB_t(\mathbf{P}_t)$$

$$= \max_{\mathbf{P}_t \in \mathbf{F}} REV_t(\mathbf{P}_t) - C_t^{BS}(\mathbf{P}_t) - C_t^{DEG}(\mathbf{P}_t) \quad (5)$$

$$= \max_{\mathbf{P}_t \in \mathbf{F}} \sum_{h \in [t+\Delta t]} \left[ \Lambda_h \left( e_h^{dis} - e_h^{cha} \right) + \Lambda_t^{rep} r_h \right] - \sum_{h \in [t+\Delta t]} c^{bs} r_h - \sum_{h \in [t+\Delta t]} \mu_t \left( e_h^{cha} + e_h^{dis} + r_h + q_t \right)$$

$$\text{s.t.} \quad E_h = (1-\rho) E_{h-1} + e_h^{cha} \eta - e_h^{dis}/\eta - r_h/\eta \quad (6)$$

$$0 \le E_h \le E^{\max} \quad (7)$$

$$0 \le r_h \le E^{\max} \quad (8)$$

$$\sum_{h \in [t+\Delta t]} r_h \le E_t^{rep} \quad (9)$$

$$0 \le e_h^{cha} \le E_h^{\max} \quad (10)$$

$$0 \le e_h^{dis} \le E_h^{\max} \quad (11)$$

$$\mathbf{P}_t = \{ e_h^{cha}, e_h^{dis}, r_h \} \quad (12)$$

Equations (5) - (12) simulate short-term BES scheduling, where the short-term objective modeled in Equation (5) is to maximize the benefits of a BES operation, which involves market revenue $REV_t$, degradation cost $C_t^{DEG}$ and battery swapping cost $C_t^{BS}$. Equations (6) and (7) guarantee that the energy output is no more than the power capacity of the BES. The Equation (8) restriction is that the energy generated by battery swapping is no more than the capacity of BES. Here we consider the status of the demand and price for battery swapping. Specifically, Equation (9) restricts the maximal daily energy swapping for BES, where $E_t^{rep}$ is determined by the demand-price curve of battery swapping. Equations (10) and (11) guarantee that the energy generated by battery charging and discharging is no more than the maximal amount of energy that can be charged/discharged in time $h$. Equation (12) indicates the decision variables involved in the model. $\mu_t$ is the AMDC for short-term scheduling, which indicates the short-term MDC of BES and can be adjusted by Equation (13), where $\mu$ is the life-cycle MDC.

$$\mu_t = \frac{\mu}{\delta_t} = \frac{\mu}{(1+r)^{-\kappa(t)}} \tag{13}$$

Additionally, to comprehensively evaluate the benefits of BES coordinated battery charging, discharging and swapping, we introduce a metric named the average benefit of usage (ABU), which is defined as follows:

$$ABU = \frac{LB^*}{D} \tag{14}$$

**2.3 Parameter setting**

In the simulation procedures, we use the decision framework and the corresponding model to simulate the life-cycle operation of a battery swapping station. In addition to energy arbitrage, we assume that battery swapping station can provide reserve as a secondary application. Specifically, we use the LMPs and non-spinning reserve prices in California Independent System Operator (CAISO) day-ahead markets in 2018 to represent the price scenarios of each year during the battery lifetime. As is common in the literature [36] [37], we use a price-taker assumption for battery swapping station in electricity markets. This means that the actions of the battery swapping station have a negligible impact on the electricity prices in the case areas. We use the battery swapping station reported in [37], which has an energy capacity of 2.7 MWh and a power capacity of 2.7 MW. Without loss of generality, we use SOH to evaluate the cumulative degradation of BES, which is defined as the ratio of the remaining energy capacity to the initial energy capacity. We assume that after 2000 charge-discharge cycles at 100% DOD, the remaining energy capacity of the BES decreases to 80% of the initial energy capacity [38], which means the physical life of the battery ends. The calendar degradation of BES is 1% capacity loss/year [39][40]. The charging/discharging efficiency is 95%, the self-discharge rate is set to 0 [4] and the labor cost of battery swapping is set to $10/h [37]. We set the discounting rate to 7%, which is recommended in [3].

# 3 Results

**3.1 Life-cycle revenue**

In the case study, we use the proposed framework and model to evaluate the life-cycle revenue of a battery swapping station and present the results in Fig. 2. Concretely, we present the arbitrage revenue, ancillary service revenue, life-cycle revenue (sum of the arbitrage revenue and ancillary service revenue) and the optimal MDC. In the above case, the life-cycle revenue increases drastically as the MDC increases from 0 to $35/MWh-throughput and decreases sharply after $35/MWh-throughput. Therefore, the battery swapping station has the highest life-cycle revenue when the MDC is $35/MWh-throughput. Additionally, the revenue of battery swapping station mainly comes from ancillary service markets when the MDC is greater than $70/MWh-throughput, which indicates that it is uneconomical to use battery swapping station for energy arbitrage when the MDC is too high. That is, in the later part of the battery's life, the benefits of battery swapping station are mainly derived from ancillary services rather than from energy arbitrage.

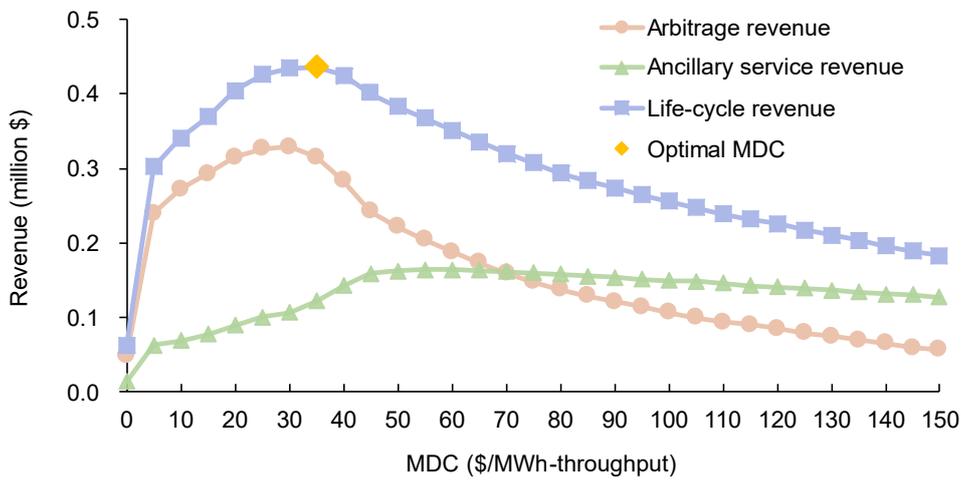

Fig. 2 Life-cycle revenues for a battery swapping station

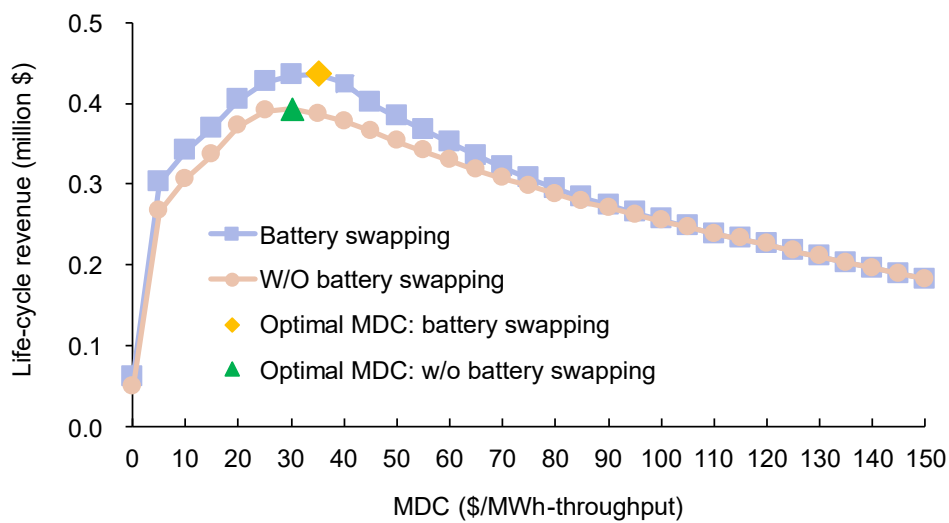

Fig. 3 Life-cycle revenues for different operation models of BES

We also report the life-cycle revenue for different BES operation models (with/without battery swapping) in Fig. 3. In Fig. 3, we can observe that the optimal MDC of the BES without battery swapping is $30/MWh-throughput, which is considerably less than that of the BES with battery swapping. Note that BES with battery swapping is substantially more profitable than that without battery swapping when the MDC is between 10 and $80/MWh-throughput. Therefore, we can use the proposed decision framework and model to choose the optimal operation modes for BES given the price scenarios and BES profiles.

**3.2 Battery life and SOH**

Here we analyze the battery lifetimes with different MDCs. As shown in Fig. 4, the battery life of the BES with battery swapping always ends earlier than that without battery swapping when the MDCs are relatively low (i.e., 0 to $70/MWh-throughput). The above results can be accounted for by the battery swapping station's substantial arbitrage capability. Energy arbitrage requires a charge-discharge process with a considerable LMP difference. Without loss of generality, in the case study, a BES with battery charging can continue to charge and discharge for up to one hour, which would lead to a limited energy arbitrage capacity. However, BES with battery swapping can cope well with the above problem. Specifically, battery swapping enables a continuous charging duration for a period longer than the maximum charging duration (one hour in this case), which considerably increases the battery asset utilization efficiency.

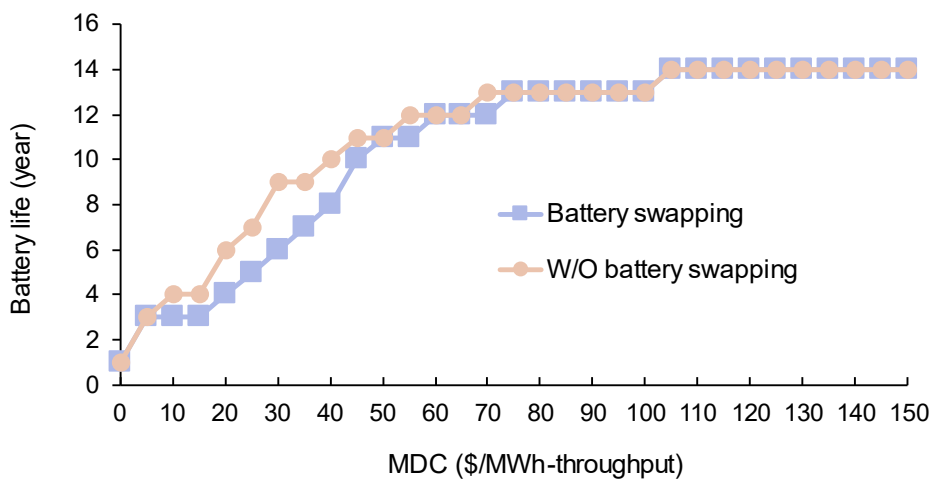

Fig. 4 Battery life for different operation models of BES

Additionally, we report the battery SOH of BES with different MDCs in Fig. 5. In Fig. 5, we can observe that the larger the MDC, the slower the SOH decreases because the MDC is defined as an opportunity cost to characterize the future long-term value of BES. That is, the future benefits of BES will be weighed higher if the life-cycle MDC is relatively large. The optimal life-cycle MDC achieves a reasonable trade-off between the utilization and battery life.

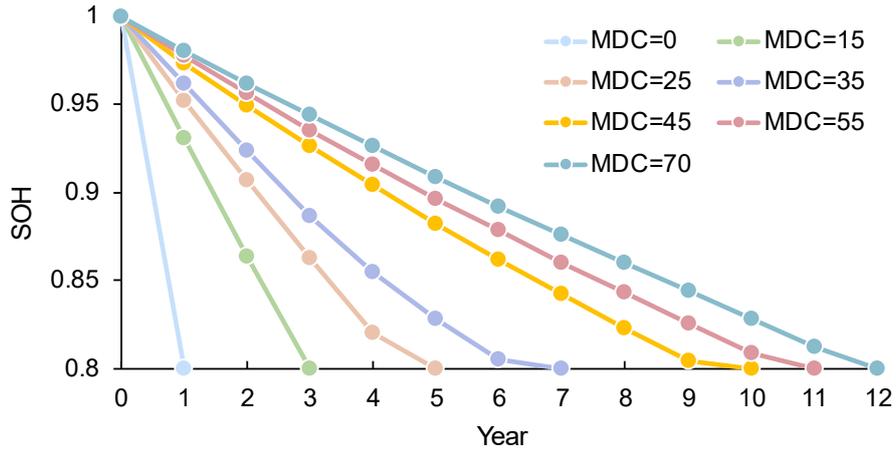

Fig. 5 Battery SOH of BES with different MDCs

**3.3 Economic and physical life**

The concept of economic end of life (EOL) for BES is proposed in [23], which is defined as the point in time beyond which the BES is unable to earn positive net profit through continued operation. To assess the economic and physical EOL of BES, the annual fixed O&M cost is incorporated into the life-cycle decision model. The annual fixed O&M cost of BES is estimated to be between 0 and $30/kW, depending on the region and project of the BES [23][41]. Here we show the cash flows of a commercial/industrial BES with an annual fixed O&M cost of $16/kW [42][43] in Fig. 6. In Fig. 6, we can observe that the net annual profit of BES decreases each year, and the negative profit occurs in the seventh year. That is, the revenue of BES cannot compensate for the fixed O&M cost in the seventh year; therefore, the economic EOL of BES is Year 6. However, the physical EOL of BES is Year 7, which indicates that the economic EOL of BES occurs faster than the physical EOL. That is, although BES can continue to operate in the seventh year, it cannot earn net profit from energy arbitrage in power and transportation systems.

The annual fixed O&M cost may be substantially different because of the various regions, markets, and BES projects. To comprehensively evaluate the economic and physical life of BES with different operation models, we conduct a sensitivity analysis on the annual fixed O&M cost of BES, which is reported in Fig. 7. As shown in Fig. 7, the physical EOL of BES with battery swapping can occur considerably faster than that of BES without battery swapping, while the economic EOL of the former occurs notably later than that of the latter. Note that in power and transportation systems, the economic life of BES depends on the arbitrage capability, which means that BES with battery swapping is more profitable than BES without battery swapping (although battery swapping does shorten the physical life of BES).

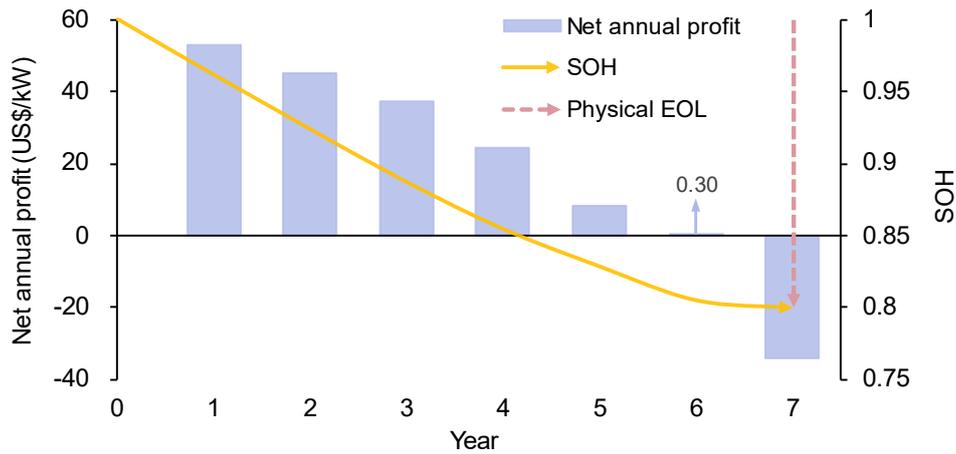

Fig. 6 Cash flows of BES in California with a fixed O&M costs $16/kW

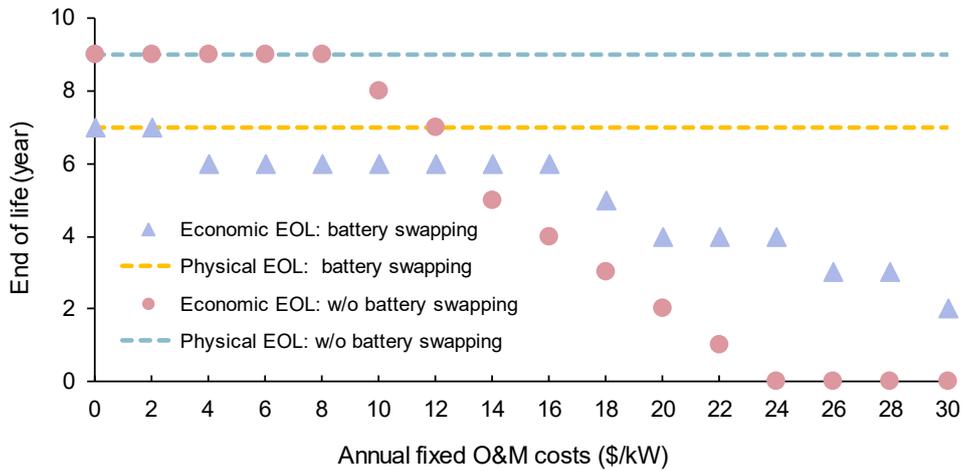

Fig. 7 Economic and physical EOL for different operation models of BES

### 3.4 Battery swapping price optimization

To analyze how the battery swapping price changes the optimal life-cycle revenue, ABU, MDC and the corresponding battery life and further determine the optimal battery swapping price, in this section, we report the life-cycle operating results of a BES with battery swapping prices from 0 and $200 in Fig. 8 and Fig. 9. Fig. 8 shows the marginal revenue of the battery swapping price in the BES operation, where the full line with rectangle marks and the green dotted line with an arrow represent the life-cycle revenue and ABU of BES with battery swapping, respectively, while the full line with cycle marks and the yellow dotted line with an arrow indicate the life-cycle revenue and ABU of BES without battery swapping.

Fig. 8 shows that the life-cycle revenue and ABU of BES with battery swapping increase considerably when the battery swapping price increases from $80/MWh to $160/MWh and decreases sharply when the battery swapping price is more than $170/MWh. The ABUs of BES with battery swapping are more than that of BES without battery swapping even though in the extreme case where battery swapping is free (ABU of BES with battery swapping is $37/MWh while that of BES without battery swapping is $36/MWh when battery swapping price is 0) also effectively validates that BES with battery swapping is more profitable than that without battery swapping. Note that the benefits of battery swapping will not increase indefinitely with an increase in the battery swapping price because the demand for battery swapping changes in the opposite direction with the price (we use the maximal amount of energy that can be swapped to characterize the demand for battery swapping in the decision model). That is, an increase in the battery swapping price will inhibit the demand for battery swapping.

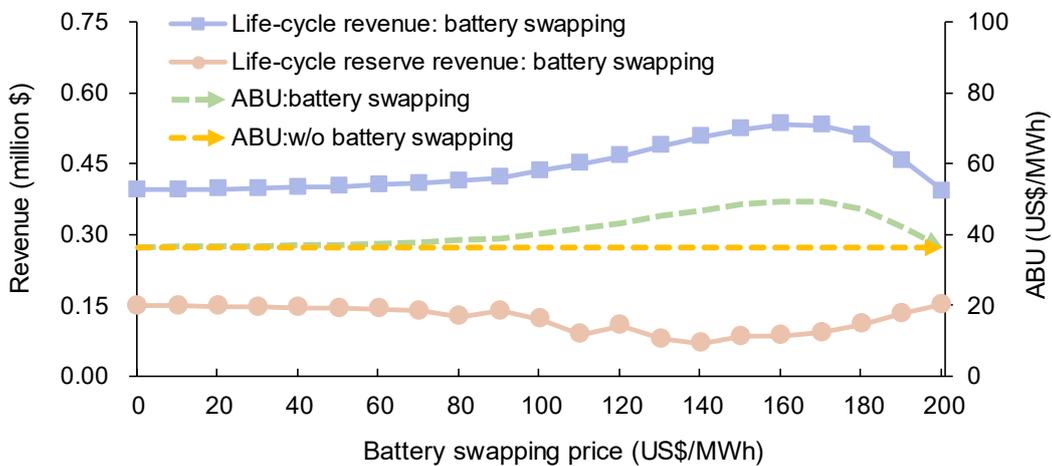

Fig. 8 Optimal life-cycle revenue and the corresponding ABU of BES for different battery swapping prices

Fig. 9 reports the optimal MDC and the corresponding lifetime of BES given the battery swapping price. Likewise, Fig. 9 shows that the optimal MDC of BES increases from $30/MWh-throughput to $45/MWh-throughput when the battery swapping price increases from $80/MWh to $160/MWh, while the corresponding battery lifetime tends to decrease with an increasing battery swapping price. The optimal MDC decreases from $45/MWh-throughput to $30/MWh-throughput when the battery swapping price increases from $160/MWh to $200/MWh, while the corresponding battery lifetime increases remarkably with increasing battery swapping price. Additionally, together with Fig. 8, we can observe that, within the price-taker assumption, a relatively high price would lead to frequent battery swapping to complete the charge-discharge cycles quickly for energy arbitrage, which would inevitably accelerate the battery's end of life. However, the above situation will not continue indefinitely with an increase in the battery swapping price because of demand-price theory.

Concretely, in the case study, the demand for battery swapping tends to be zero when the battery swapping price is more than $180/MWh. In this sense, EVs only charge or discharge in BES instead of swapping batteries, which would lead to an increase in battery lifetime and a decrease in the optimal MDC.

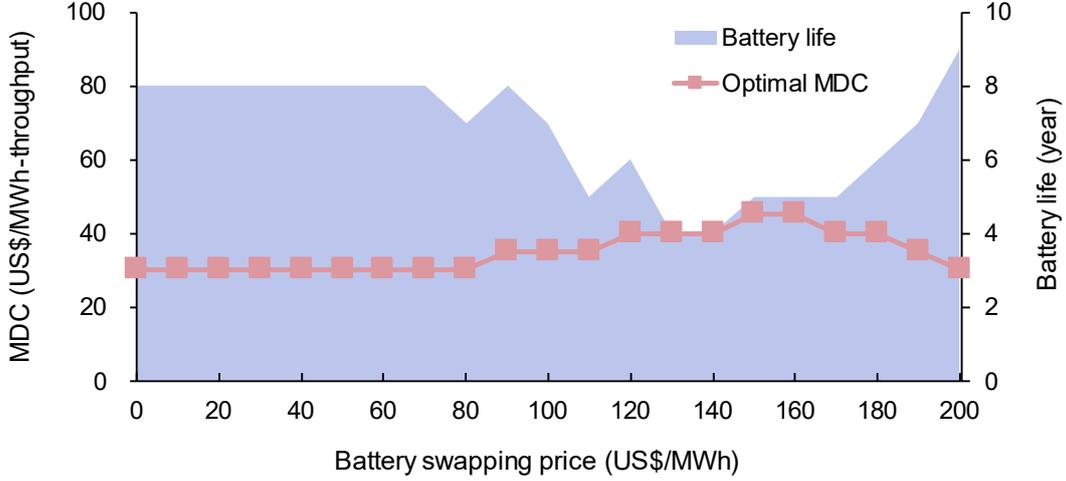

Fig. 9 Optimal MDC and the corresponding lifetime of BES for different battery swapping prices

Here we also consider the price elasticity of the demand for battery swapping and analyze the optimal battery swapping price strategy for different demand-price curves. Conventionally, a daily demand-price curve of battery swapping can be formulated in linear form $\Lambda_t^{rep} = kE_t^{rep} + b$, where $\Lambda_t^{rep}$ and $E_t^{rep}$ represent the daily battery swapping price and the amount of battery swapping, respectively, and $k$ and $b$ indicate the slope and vertical intercept of the demand-price curve, respectively. Specifically, we use 10 demand-price curves, which are shown in Fig. 10, where each demand-price curve represents a predictable market and the combinations of $(k,b)$ indicate the sensitivity of the demand-price (the price elasticity of the demand essentially reflects the sensitivity of the demand to price changes). Together with Fig. 10, the sensitivity of the demand price varies dramatically in different curves.

In this context, we also report the test results of the price elasticity of the demand for battery swapping in Fig. 10, where the scatter points represent the optimal battery swapping price and the corresponding battery swapping demand for different demand-price curves. In Fig. 10, we can observe that the optimal price and demand of battery swapping vary substantially in markets with different demand-price curves. Note that the demand for battery swapping tends to be zero or even negative when the battery swapping price is relatively high, which indicates that EVs only charge and discharge in BES instead of swapping batteries in the current market. Without loss of generality, the greater the price elasticity of the demand of a market, the smaller the optimal battery swapping price. That is, a slight increase in the battery swapping price would lead to a sharp decrease in the demand for battery swapping in a market with a relatively high price elasticity of demand, which would further account for the lower revenue of a battery swapping station.

In summary, the test results of the price elasticity of the demand can provide decision support for the owners of battery swapping stations to determine the optimal pricing strategy of battery swapping and can further estimate the corresponding life-cycle ABU and MDC by the intertemporal decision framework given a predictable market.

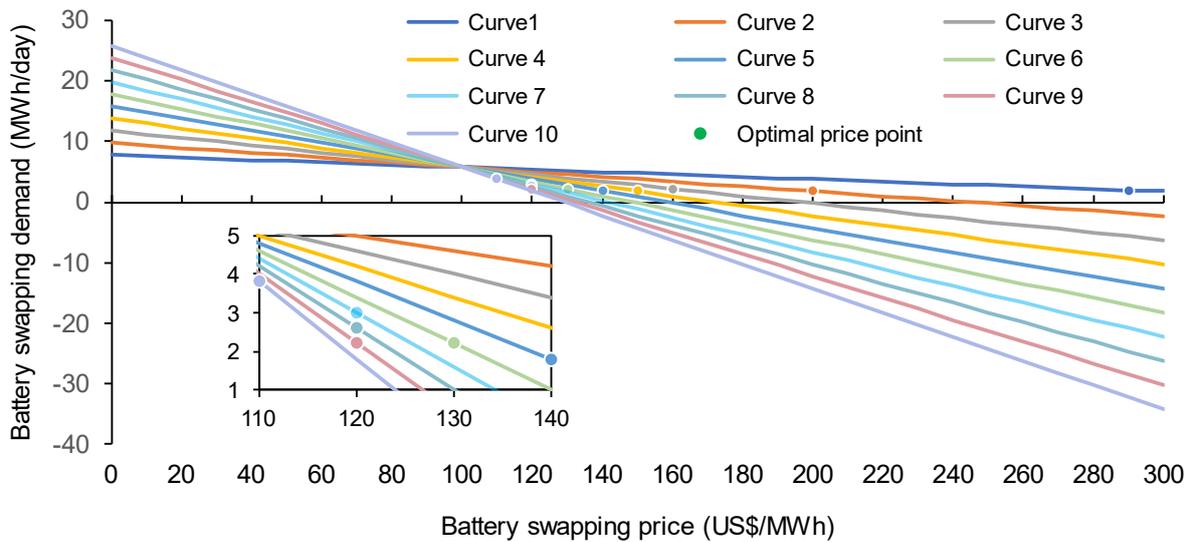

Fig. 10 Optimal demand and price of battery swapping for different demand-price curves

## 4 Discussion

Although the BES with battery swapping is more profitable than the BES without battery swapping, the physical EOL of the BES with battery swapping occurs considerably earlier than that of the BES without battery swapping, which can be accounted for by the frequency of battery use caused by battery swapping. Intuitively, we can extend the operating life of BES with battery swapping by adding new battery packs. However, it may be uneconomical because additional capital costs for adding new battery packs are needed. Moreover, the net profit earned from BES with battery swapping compared with BES without battery swapping may not compensate for the capital costs of the new battery packs. Therefore, in the life-cycle operation period, we need to make a reasonable trade-off between benefits and battery life for a BES with battery swapping. Specifically, we can control the utilization and benefit of BES by adjusting the life-cycle MDC when planning a project with a specific business model or lifetime requirement.

Furthermore, although the economic EOL of BES with battery swapping usually occurs later than that of BES without battery swapping, the economic EOL of BES with battery swapping still occurs considerably earlier than its physical EOL when the fixed O&M costs are relatively high. That is, BES cannot earn positive net profit through continuous operation in the above situation. In this context, BES is recommended for use in the second-hand market or to provide services that require less cycling capability, such as ancillary service of power grids involving contingency reserve, back-up and black-start sources. In some cases, the revenue earned by the second-hand and ancillary service markets represents a high percentage of the total revenue of BES.

Additionally, the optimal battery swapping price determined by the improved intertemporal decision framework in markets with different price elasticity of demand can provide decision support for the planning and operation of BES. Specifically, given the predictable market and BES profiles, the owners and operators of BES can be informed by the decision framework of how to determine the optimal battery swapping price and estimate the corresponding life-cycle revenue, MDC and battery lifetime of BES.

## 5 Conclusion

In this paper, we incorporate battery charging, discharging and swapping into the decision framework and BES model. Specifically, we estimate the optimal life-cycle benefits and determine the optimal MDC, show the economic and physical EOL of BES, and analyze how the battery swapping price affects the life-cycle revenue, optimal MDC and corresponding battery life using the proposed framework and model.

The operational decisions for a large-scale battery swapping station including multiple battery packs should determine whether to use, how much, or which batteries to use for different periods. Additionally, factors, such as the regions of battery swapping stations and market fluctuations, may change battery swapping prices. Future works will be focused on the decision model and economic evaluation framework for a large-scale BES.

## 6 Acknowledgments

This work was supported in part by the National Key Research and Development Program under Grant 2022YFB2405600 and in part by the National Natural Science Foundation of China under Grants 72271008, 72131001, T2121002, and 52277092.